\documentclass[preprint,prd,noshowpacs]{revtex4}
\usepackage{amssymb}
\usepackage{amsmath}
\usepackage{graphicx}  

\begin{document}

\title{Gravity's light in the shadow of the Moon
\footnote{Essay written for the Gravity Research Foundation 2018 Awards for Essays on Gravitation. \\ Awarded ``Honorable Mention" \\}}

\author{Andri Gretarsson}
\thanks{Andri.Gretarsson@erau.edu}
\author{Preston Jones}
\thanks{Preston.Jones1@erau.edu}
\affiliation{Embry Riddle Aeronautical University, Prescott, AZ 86301}
\author{Douglas Singleton (corresponding author)}
\thanks{dougs@mail.fresnostate.edu}
\affiliation{California State University Fresno, Fresno, CA 93740}

\date{\today}

\begin{abstract}
In this essay we look at the possibility of vacuum production of very low frequency electromagnetic radiation from a gravitational wave background ({\it i.e.} gravity's light). We also propose that this counterpart electromagnetic radiation should be detectable by a lunar orbiting satellite which is periodically occulted by the Moon ({\it i.e.} in the shadow of the Moon). For concreteness we consider the possibility of detection of both the gravitational wave and hypothesized electromagnetic radiation counterpart from the supernova core collapse of Betelgeuse.            
\end{abstract}

\maketitle

{\noindent {\large {\bf Gravity's light...}}}

The earliest effort to demonstrate the relation between gravity and electricity was reported by Faraday in 1850 \cite{Faraday1885}, ``Here end my trials for the present. The results are negative. They do not shake my strong feeling of the existence of a relation between gravity and electricity, though they give no proof that such a relation exists." Faraday conducted experiments to measure gravitoelectric induction from the Earth's gravitational field and after eliminating any geomagnetic induction was unable to measure any relation between gravity and electricity. More recently Skobelev \cite{Skobelev75} established (theoretically) the link between gravity and electromagnetism by calculating the tree-level Feynman diagrams for the process $graviton + graviton \to photon + photon$. Although this process was found to be non-zero it was a very small effect. Skobelev found that the cross-section for two gravitons to annihilate into two photons was on the order of $10^{-110} ~ \rm{cm^2}$ for photons having an energy/frequency corresponding to the rest mass of the electron. 

We recently revisited the theoretical work of Skobelev by considering an incoming gravitational plane wave and an outgoing electromagnetic plane wave created from the gravitational wave  \cite{Jones15,Jones16,Jones17}. This process of an incoming gravitational plane wave creating an outgoing electromagnetic plane wave can  be understood as the semi-classical production of photons from the vacuum in the ``external field" of the gravitational wave background, {\it i.e.} the interactions of a collection of gravitons and photons rather than the individual quanta. To find the production of electromagnetic radiation from an incident gravitational wave we took the source free electromagnetic field Lagrange density in the Lorenz gauge as $\mathcal{L}_{em}  =  - \frac{1}{2}\partial _\mu  A_\nu  \partial ^\mu  A^\nu$. Separating the vector potential as $A_\mu  \left( {\kappa,\lambda ,x} \right) = \epsilon_{\mu} ^{(\lambda)} \phi ^{(\lambda)} \left( {\kappa ,x} \right)$ results in a scalar-like Lagrange density $\mathcal{L}_{em}  =  - \partial _\mu  \varphi ^* \partial ^\mu  \varphi$, where $\varphi = \frac{1}{\sqrt{2}} \left( \phi^{(1)} + i \phi^{(2)} \right) $. The field equations for the scalar field, $\varphi (x)$, in a general gravitational background is

\begin{equation}
\frac{1}{{\sqrt{-g}}} \partial_{\mu} \sqrt{-g} g^{\mu \nu} \partial_{\nu}\varphi = 0.
\label{eomvarphi}
\end{equation}

\noindent For the metric we take a plane gravitational wave background of the form

\begin{equation}
ds^2 = -dt^2 + dz^2 + a(u)^2 dx^2 + b(u)^2 dy^2, 
\label{GWmetric}
\end{equation}

\noindent where $u=z-t$ is one of the standard light front coordinates with $c=1$. We also take the metric to oscillate sinusoidally, which means $a(u) =1 + \varepsilon (u) $ and $b(u) = 1 - \varepsilon(u)$ with $\varepsilon(u) =h e^{iku}$, $h$ is a strain amplitude and $k$ the wave number of the gravitational wave. With the metric \eqref{GWmetric} the general solution  \cite{Jones16} to \eqref{eomvarphi} is

\begin{equation}
\varphi  = A e^{\frac{\lambda }{k}} e^{ - \frac{\lambda }{{k\left( {1 - h ^2 e^{2iku} } \right)}}} \left( {1 - h ^2 e^{2iku} } \right)^{\frac{1}{2}\left( {\frac{\lambda }{k} - 1} \right)} 
e^{ - i\lambda u} e^{ip_v v} e^{ip x} e^{ip y} + B ~,
\label{Sfield}
\end{equation}

\noindent where $A$ and $B$ are arbitrary constants, $\lambda =\frac{p^2}{2 p_v}$  and $p, p_v$ are $\varphi$ field momenta in the $x,y$ and $v=z+t$ directions. Choosing the constants in (\ref{Sfield}) as $B=-A=-1$ and taking the field momenta to zero ($p, p_v, \lambda \to 0$), as the vacuum limit of the scalar field, we find that the outgoing wave solution for $\varphi$ does not vanish,  

\begin{equation}
\varphi \left( {t,z} \right) = \left( {1 - h^2 e^{2ik\left( {z - t} \right)} } \right)^{ - \frac{1}{2}} -1  \approx \frac{1}{2} h ^2 e^{2i k (z-t)}+ \frac{3}{8}  h ^4 e^{4i k(z-t)},
\label{OutState}
\end{equation}

\noindent where in the last expression we have taken $h$ to be small. 

The result in \eqref{OutState} and the non-zero amplitude calculated for $graviton + graviton \to photon + photon$ in \cite{Skobelev75} appear to be at odds with the well known prohibition on particle/field production by a gravitational wave \cite{gibbons}. However, the prohibition on production has a caveat, ``unless the created particles were massless and precisely aligned with the momentum of the plane wave spacetimes". The solutions \eqref{OutState} precisely satisfy these exceptions -- the field $\varphi$ is massless and the scalar field plane wave travel in the same direction as the incident gravitational plane wave. More concretely one can calculate the Bogoliubov coefficients, $\beta_{ij}$, for the above situation of a field and a plane wave gravitational background, where the Bogoliubov coefficients are a measure of field production. Reference \cite{garriga} calculated these Bogoliubov coefficients and found  $\beta_{ij} = \langle u_i ^{out} | u_j ^{in~*} \rangle \propto \delta (k_{-} + l_{-})$ with $u_i ^{out} , u_j ^{in}$ being the outgoing and incoming scalar field modes, and $k_{-} , l_{-}$ are the light front coordinate momenta of $u_j ^{in} , u_i ^{out}$ respectively. For a massive scalar field $k_{-} + l_{-} \ne 0$, the argument of the Dirac delta is never zero, and $\beta_{ij} = 0$. However for a massless field traveling in the same direction as the gravitational wave background $k_{-} + l_{-} \to 0$, the argument of the Dirac delta does go to zero, and $\beta_{ij} \ne 0$. This is ``gravity's light" the electromagnetic radiation produced by the gravitational wave background.

{\noindent {\large {\bf ... in the shadow of the Moon}}}

Having found theoretical support for Faraday's ``strong feeling of the existence of a relation between gravity and electricity", it remains to determine how one might observe this ``gravity's light". The first question that must be answered is ``How strong is the electromagnetic radiation generated from a given gravitational wave?". This question was addressed in our previous work \cite{Jones17} using the Newman-Penrose scalar invariant formalism \cite{NP}. First, the electromagnetic invariant $\Phi_2$ was found using $\varphi$ in \eqref{OutState}, and the gravitational invariant $\Psi _4$ was found using the plane wave metric of \eqref{GWmetric}. In terms of these invariants the general ratio of electromagnetic to gravitational energy flux was given by, $\frac{d E_{em}}{d E_{gw}} = \frac{{\left( {\frac{1}{{4\pi }}\left| {\Phi _2 } \right|^2 } \right)}}{{\left( {\frac{1}{{16\pi k^2 }}\left| {\Psi _4 } \right|^2 } \right)}} $. Substituting the values of $\Phi _2$ and $\Psi _4$ from \eqref{GWmetric} and \eqref{OutState} in the energy ratio gives a flux ratio of $F_{em}   = 2 h^2 F_{gw}$. The flux for the gravitational plane wave, $F_{gw}$, can be calculated from the metric in \eqref{GWmetric} yielding $F_{gw} ^{(0)} = \frac{{c^3 }}{{16\pi G}}\left| {\dot \varepsilon} \right|^2  = \frac{{c^3 h_0^2 \omega^2}}{{16\pi G}}$, where $\varepsilon (u) = h e^{iku}$ from the definition of the metric ansatz functions, $a(u), b(u)$. Finally, substituting this into the flux ratio gives

\begin{equation}
F_{em} ^{(0)} =  \frac{ \pi {c^3}}{{2 G}} h_0^4 f^2 ,
\label{emFlux}
\end{equation}

\noindent where $h_0$ is the strain amplitude at the point of production, and $f$ is the gravitational wave frequency. Using \eqref{emFlux} the electromagnetic flux at some distant observation point $R$ is

\begin{equation}
F_{em} = \left( {6 \times 10^{35}~\rm{ \frac{{Ws^2 }}{{m^2 }}}} \right) \left( \frac{r_0}{R} \right)^2  h_0^4 f^2   ~,
\label{emFlux2}
\end{equation}

\noindent where $r_0$ is the region of electromagnetic radiation production and $h_0$ is the associated strain amplitude.

One of the strongest possible local signal strengths for ``gravity's light" is the imminent (astronomically speaking) supernova of our neighbor star Betelgeuse which is at a distance of $R \sim 200$ parsecs or $6 \times 10^{18}~\rm{m}$. The first gravitational waves generated from the Betelgeuse core collapse supernova, that could be detected at Earth, would be from the core bounce and the subsequent ``convective mass flow" \cite{Muller04}. The maximum expected quadrupole amplitudes following the core collapse are between $10~\rm{cm} - 100~\rm{cm}$, where the quadrupole amplitude is the constant ratio of strain amplitude to inverse distance from the source. The associated gravitational wave strain amplitudes at the Earth are $10^{-19} - 10^{-20}$ at a nominal frequency of $f \sim 1~\rm{kHz}$. Such a signal is well within the current sensitivity curves of gravitational wave detectors. This prompt gravitational radiation from core collapse would have a strain amplitude, just outside the collapsing star at $r_0 \sim 10^{12}~ \rm{m}$, of $h_0 \sim 10^{-13}$ from the bounce, and $h_0 \sim 10^{-12}$ from the convective mass flow. Collecting these values in \eqref{emFlux2} the associated vacuum produced electromagnetic flux at the Earth is $10^{-24} ~\rm{ \frac{{W }}{{m^2 }}}$ from core bounce and $10^{-20} ~\rm{ \frac{{W }}{{m^2 }}}$ from convective mass flow. These are strong signals and should be readily detectable with modest equipment.

The reader might find themselves thinking ``this seems too easy", and indeed it is not quite that easy. The interstellar medium has a plasma cutoff for electromagnetic radiation of $1-3$ kHz {\it i.e.} electromagnetic waves below $1-3$ kHz will be exponentially attenuated. Since the characteristic frequency of the gravitational waves considered above is $f \sim 1$ kHz the counterpart electromagnetic radiation would be about $f_{em} \sim 2$ kHz. This can be seen from the solution for $\varphi$ which has oscillation terms like $e^{2iku}$ whereas the gravitational wave has $e^{iku}$. Despite this doubling of the frequency, the electromagnetic radiation produced from the gravitational wave will still be VLF, comparable to the interstellar plasma cutoff frequency, and at ``the end of the rainbow" \cite{Lacki10} for observation. Such VLF electromagnetic radiation, with $f_{em} \sim 2$ kHz, would be highly attenuated as it traveled into the inner solar system. The plasma cutoff frequency grows roughly linearly as one travels toward the inner Solar System, and 2 kHz is well below the local plasma frequency cutoff. It is only in the outer heliosphere that the plasma frequency cutoff is sufficiently low to have any chance of detecting VLF radiation in the frequency range of $f_{em} \sim 2$ kHz. In fact the Voyager spacecraft, which are in the outer regions of  the heliosphere, have detected VLF signals just above a few kHz, using plasma density detectors. However these signals are generally believed to be associated with Solar activity \cite{Kurth84,Kurth03}. In any case the prospect of detecting the gravitational wave produced electromagnetic signals of a few kHz is not very promising unless one wants to send detectors, like the two Voyager spacecraft, to the outer heliosphere where the plasma cutoff frequency is sufficiently low.  

To observe the VLF counterpart electromagnetic radiation nearer to the Earth than the outer heliosphere one must consider higher frequency gravitational waves to avoid the $\sim 20 - 30$ kHz plasma cutoff frequency near the Earth. Recalling that the VLF electromagnetic radiation has twice the frequency of the gravitational wave means one would need gravitational waves with a frequency range of $10-15$ kHz or greater. Following the core collapse and supernova, Betelgeuse is generally expected to form a proto-neutron star and ultimately a neutron star. The highly energetic proto-neutron star and later neutron star would have star-quakes which will produce gravitational radiation from several modes at relatively high frequencies \cite{Sontani17}. Two of these modes are the $f$-modes (for fundamental mode) and the $w$-modes (for wave mode). The $f$-modes are expected to produce a stronger signal but with a frequency of $1-3$ kHz, while the weaker $w$-modes have a frequency range of $8-16$ kHz. Doubling the gravitational wave $w$-mode frequency would give a range of $16-32$ kHz for the hypothesized counterpart VLF electromagnetic radiation. The frequency range for the VLF radiation from $w$-modes is above the plasma cut-off frequency in the inner Solar System and this ``gravity's light'' could be detected near the Earth. The counterpart electromagnetic radiation for the lowest order $w$-mode from the proto-neutron star should be detectable at the Earth {\it if} one could shield the interference from the Sun. To shield the Sun's interference, a satellite could be placed in a lunar orbit periodically occulted by the Moon similar to the old Explorer 49 \cite{Jones17} and in an orbit optimized for observation of VLF radiation. Thus we propose searching for ``gravity's light in the shadow of the Moon" using an updated version of the old Explorer 49 satellite. With a little patience, after the ``dust settles" from a Betelgeuse supernova, gravitational waves from the neutron star-quake $w$-modes would produce counterpart electromagnetic radiation which could be detected by a lunar orbiting, and periodically occulted, satellite. 

Betelgeuse is certainly not the only star in the Milky Way Galaxy with the potential to go supernova and form a neutron star, but these events are extremely rare. Fortunately, the planned lunar occulted detection is rendered much more practical by the greatly higher event rates of Galactic neutron star quakes that could be routinely detected. Detection of the counterpart radiation produced by star quake gravitational waves would add a completely new tool to the current multi-messenger model of astrophysics. However, the real importance of the proposed experiment was best described by Faraday \cite{Faraday1885}, ``Such results, if possible, could only be exceedingly small; but, if possible, i.e. if true, no terms could exaggerate the value of the relation they would establish.".

{\noindent {\bf Acknowledgment:}} We would like to thank Darrel Smith for bringing Faraday's work on gravity and electricity to our attention.  PJ was partially supported by the Embry Riddle Aeronautical University Faculty Research Development Program. 


\end{document}